\documentclass[aps,prl,twocolumn]{revtex4}

\usepackage{amsmath,amssymb} 
\usepackage{bm} %
 
\usepackage{color} %
\usepackage{graphicx}

\usepackage[colorlinks]{hyperref} %
\hypersetup{
citecolor=blue,
linkcolor=blue,
urlcolor=blue
}

\usepackage{epstopdf}
\usepackage{comment}

\usepackage{multirow}

\newcommand{\fig}[1]{Fig.~\ref{#1}}

\newcommand{\tab}[1]{Tab.~\ref{#1}}

\newcommand{\eq}[1]{Eq.~(\ref{#1})}

\newcommand{\sigmav}{\langle \sigma v \rangle}

\newcommand{\pbar}{\bar p}

\begin{document}
\title{Antiprotons from dark matter annihilation through light mediators
and \\ a possible excess in AMS-02 $\bar{p}/p$ data
}
\author{Xian-Jun Huang}
\author{Chun-Cheng Wei}
\author{Yue-Liang Wu}
\author{Wei-Hong Zhang}
\author{Yu-Feng Zhou}
\affiliation{
	CAS Key Laboratory of Theoretical Physics, 
	Institute of Theoretical Physics, Chinese Academy of Sciences,
	ZhongGuanCun East Rd.55, Beijing, 100190, China.
	}
\affiliation{
	University of Chinese Academy of Sciences, 
	Yuquan Rd.19A, Beijing 100049, China
}
\begin{abstract}
We show that in the scenario where 
dark matter (DM)  particles annihilate 
through light mediators, 
the energy spectra of the final state cosmic-ray particles  depend strongly  on 
the mediator mass.
For  final state antiprotons, 
a spectrum with  relatively narrow peak occurs when
the mediator mass is comparable to the  $\bar{p} p$ production threshold.
Of interest, the latest  AMS-02 data on the $\bar{p}/p$ flux ratio 
hint at a bump-like excess over the expected background
in the energy range $\sim100-450$ GeV.
We show that such a light mediator scenario is favoured  by 
the latest AMS-02 data over the scenarios of 
DM direct annihilation into the standard model particles 
and 
that of antiprotons produced from inside supernova remnants (SNRs), 
and is  consistent with the upper limits derived from 
the Fermi-LAT data on
the gamma rays towards the dwarf spheroidal galaxies.
The  $\bar{p}/p$ flux ratio with energy above 450 GeV  is predicted to 
fall with energy quickly, 
which  can be easily distinguished  from
the other two scenarios
as they  predict  the  ratio to be flattening or rising up to multi-TeV region.
\end{abstract}
\pacs{xx.xx}
\preprint{\tt [\today ]}
\maketitle %

\paragraph*{Introduction.}
Cosmic-ray (CR) antiparticles such as positrons and antiprotons
are relatively rare and sensitive to exotic contributions.
In  recent years, 
an excess over the expected background in 
CR positrons has been observed 
\cite{positronFrac}.
The spectral feature of the excess plays an important role 
in identifying  its origin  such as nearby  astrophysical sources or
dark matter (DM) interactions.
Recently, 
the AMS-02 collaboration published the  measurement on 
the antiproton to proton ($\bar{p}/p$) flux ratio   up to 
kinetic energy $450$~GeV,
based on four years of data taking
\cite{Aguilar:2016kjl}, %
confirming the first preliminary result presented in the year 2015~\cite{Ting:AMS}.
Although the measured kinetic energy spectrum of $\pbar/p$ ratio 
is in overall agreement with the secondary production 
especially below $\sim 100$~GeV
\cite{
Giesen:2015ufa,
Jin:2015sqa%
},
at higher energies, there is a trend of flattening and smooth rise
in the range $\sim 100-260$~GeV, 
followed by  a drop by $\sim30\%$ in the range $\sim 260-450$~GeV.
Such a hint of a bump-like excess has already been observed in 
the preliminary AMS-02 result~\cite{Ting:AMS}, and 
is strengthened in the latest data with 
higher statistics.

This intriguing possibility of an excess with 
a distinctive spectral feature in $\bar p/p$ flux ratio, 
if confirmed, %
may shed light on the nature of its origin:
i) The  pulsar wind nebulae are unlikely to  
produce  energetic  antiprotons.
ii) In the leading astrophysical explanation, 
extra antiprotons can be  produced from 
collisions of primary CRs with the gas inside 
supernova remnants (SNRs),
the resulting energy spectrum, however, features 
a continued flattening or weak rise at least up to 1 TeV
for a typical  cut-off energy 
$\mathcal{E}_{\text{max}}\sim \mathcal{O}(10)$~TeV
\cite{
Blasi:2009bd,%
Mertsch:2014poa%
}.
iii) The spectrum of antiprotons produced from 
halo DM annihilation directly into standard model (SM) final states 
in general features a very broad bump
due to the long chain of cascade showers and 
hadronization of the final state partons.
Since there is little room left for  extra contributions below $\sim100$~GeV, 
the DM particle mass $m_{\chi}$ is pushed to be  very high ($m_{\chi}\gtrsim$ 2~TeV)
\cite{
Giesen:2015ufa,%
Jin:2015sqa,%
Lin:2015taa,%
Ibe:2015tma%
}.
Consequently,  only the  spectral tail of DM produced antiprotons can extend to 
the energy range accessible to 
the current AMS-02 experiment ($E\lesssim 450$ GeV),
thus again a flattening or weak rise of $\bar p/p$ ratio is expected in this region.

In this letter, we show that
in a class of  scenarios where 
DM particles annihilate  through light color-singlet mediators, 
the energy spectrum of final state particle can be  a narrow bump
with  reduced multiplicity.
For antiprotons, a  narrow peak   occurs when
the mediator mass is comparable to the $\pbar p$ production threshold $2m_{p}$.
We show that such a light mediator scenario is favoured  by the latest AMS-02 data
over the  scenario of DM direct annihilation  and 
that of antiprotons produced from inside of  SNRs, 
and is  also consistent with the known constraints such as
the gamma-ray limits from the dwarf spheroidal galaxies (dSphs). 
The  $\pbar/p$ ratio in the high energy range  is predicted to fall
with energy quickly, which makes it highly distinguishable from
the other two scenarios. 

\paragraph*{Effects of mediators.}
The annihilation of DM particles provides 
an extra primary source of  CR particles.
The corresponding primary source term for a final state particle $f$
takes the  form
\begin{align}\label{eq:ann-source}
q(\boldsymbol{r},E)=\frac{\rho(\boldsymbol{r})^2}{2 m_{\chi}^2}\langle \sigma v \rangle 
\frac{dN}{dE} ,
\end{align}
where $\langle \sigma v \rangle$ is 
the velocity-averaged DM annihilation cross section multiplied by DM relative velocity,
$\rho(\boldsymbol{r})$ is the DM energy density distribution function,
$dN/dE$ is the spectrum  of kinetic energy $E$
which is related to the total energy $\mathcal{E}$ as
$E=\mathcal{E}-m_{f}$
with  $m_{f}$ the mass of the final state particle $f$.

In many well-motivated DM models, 
the DM particles do not  couple to the SM particles directly, 
but through some light color-singlet mediator particles, 
which has rich phenomenological consequences
in DM annihilation
\cite{
sommefeld,%
Liu:2013vha,Chen:2013bi},
self-scatterings 
\cite{
Loeb:2010gj,%
Tulin:2012wi,%
Kaplinghat:2015gha%
}
and solar capture
\cite{
Chen:2015uha,%
Liang:2016yjf%
},
and direct detections
\cite{
Li:2014vza,%
DelNobile:2015uua%
}.
In this scenario, for the same DM mass,
the resulting  energy spectrum of final state particle from 
DM annihilation can be significantly different from 
the case without a mediator.
Let us consider 
DM annihilating first into two mediators $\bar\chi \chi\to 2\varphi$ and
followed by the decay $\varphi\to f +X$,
where 
$X$ stands for any other final states.
The spectrum  $dN(x)/dx$ of the scaled total energy 
 $x=\mathcal{E}/m_{\chi}$ in the DM center-of-mass frame 
is related to that in the mediator rest-frame 
$dN(x')/dx'$ (with $x'=2 \mathcal{E}'/m_{\varphi}$) by a Lorentz boost 
\begin{align}\label{eq:boost}
\frac{dN(x)}{dx}=
2\int^{b(x)}_{a(x)} dx' \frac{1}{\sqrt{1-\varepsilon_{1}^{2}}\sqrt{x^{'2}-\varepsilon_{0}^{2}}}\frac{dN(x')}{dx'} ,
\end{align} 
where 
the two parameters $\varepsilon_{1}=m_{\varphi}/m_{\chi}$ and $\varepsilon_{0}=2m_{f}/m_{\varphi}$ characterize the mass hierarchies in the two-step cascade process.
The lower and upper limits  of the integration are
$a(x)=x_{-}$ and $b(x)=\text{min}\{1,x_{+}\}$ with 
$x_{\pm}=2(x\pm\sqrt{(1-\varepsilon_{1}^{2})
(x^{2}-\varepsilon_{1}^{2}\varepsilon_{0}^{2}/4)})/\varepsilon_{1}^{2}$.
In the large hierarchy limit $\varepsilon_{0}\ll 1$,
\eq{eq:boost} reproduces the known result in Refs.
\cite{
Mardon:2009rc,%
Elor:2015tva%
}.

In general,  
 Lorentz boosts at random directions tend to broaden
the energy spectrum.
However, the effect of broadening is suppressed if
the velocity of $f$ in the mediator rest-frame is small.
In the simplest case where  $f$ is mono-energetic 
with energy $\bar{\mathcal{E}}'$, i.e.
$dN/\mathcal{E}'\propto \delta( \bar{\mathcal{E}}' - \mathcal{E}')$
in the mediator rest-frame,
the spectrum boosted  to the DM center-of-mass frame is  
a  box-shaped spectrum  with a  center energy $\bar{\mathcal{E}}$ and 
width $\Delta \mathcal{E}$ 
\begin{align}\label{eq:maxmin-energy}
\bar{\mathcal{E}}=\gamma_{B} \bar{\mathcal{E}}', \qquad
\Delta \mathcal{E}/\bar{\mathcal{E}}
=2   \beta_{B}  \beta' ,
\end{align}
where $\gamma_{B}=1/\varepsilon_{1}$ is the Lorentz boost factor,
$\beta_{B}=(1-\varepsilon_{1}^{2})^{1/2}$
is the boost velocity, and 
$\beta'=(1-m_{f}^{2}/\bar{\mathcal{E}}^{'2})^{1/2}$ is the velocity 
of $f$ in the mediator rest-frame.
For the decay of mediator into light quarks 
$\varphi\to \bar q q\to\pbar  +X$ $(q=u,d)$, 
the velocity of antiproton has an upper limit
$\beta' \leq (1-\varepsilon_{0}^{2})^{1/2}$
as $X$ at least contains a proton.
Thus when $\varepsilon_{0}\approx \mathcal{O}(1)$, namely, 
$m_{\varphi}$ is comparable to the $\pbar p $ production threshold $2m_{p}$, 
the value of $\beta'$ has to be  small and 
the spectrum is a narrow box in the DM halo rest-frame.
On the other hand, 
the total energy has a lower limit $\bar{\mathcal{E}}'\geq m_{p}$,
the value of  $\bar{\mathcal{E}}$ can be very large for
a large Lorentz boost factor $\gamma_{B}\gg 1$.
Thus after the boost,  the energy spectrum is a narrow box at high energy.
Since the antiproton energy spectrum from the decay 
$\varphi\to \bar q q\to\pbar  +X$ is low-energy
dominated, the Lorentz boost will push the antiprotons into high energy region
without significantly broadening the spectrum.
On the contrary, the antiproton spectrum from DM direct annihialtion
$\chi\chi\to \bar q q$ is very broad for the same DM mass, due to
the larger center-of-mass energy of the $\bar q q$ system and thus
the longer chain of cascade parton showers and hadronization.
Furthermore, 
in the light mediator scenario,
the multiplicities of the final state particles are suppressed by
the smallness of the mediator mass and can be much lower than
that in the case without  mediators.
For the hadronic decay $\varphi\to\bar q q$, 
the energy spectrum $dN/dx'$ of antiprotons is simulated  using
the Monte Carlo event generator Pythia 8.2
\cite{
Sjostrand:2007gs%
}.
The $\pbar$ multiplicity is found to be 
$N_{p}=$0.35, 1.2, 1.6 and 3.0 for 
the center-of-mass energy 
$\mathcal{E}_{\text{CM}}=10$, 50, 100 and 500 GeV, respectively.
Thus for the same DM mass, the DM annihilation through light mediators will generate 
less  antiprotons due to lower center-of-mass energy.

\begin{figure}[thb]
\includegraphics[width=0.8\columnwidth]{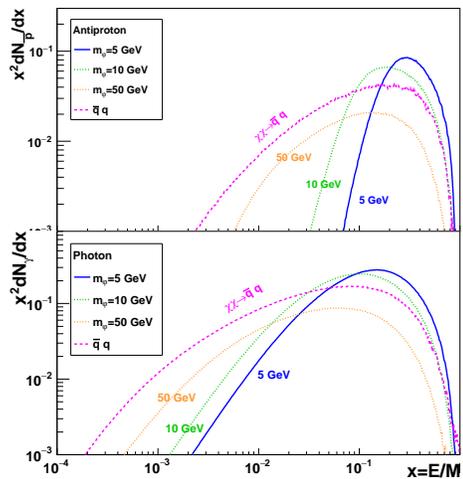}
\caption{
Upper panel)
scaled energy spectra $x^{2} dN/dx$ of antiprotons per 
DM annihilation  from the annihilation
$\chi\chi\to 2\varphi\to 2 \bar q q, \ (q=u,d)$ with mediator masses
$m_{\varphi}=5$, 10 and 50 GeV, respectively.
The DM particle mass is fixed at $m_{\chi}=500$~GeV.
The spectrum of DM direct annihilation $\chi\chi\to \bar q q$ with 
the same DM mass is also shown.
Lower panel) the same as the upper panel, but for photons. 
\label{fig:spectrum} 
}
\end{figure}

In \fig{fig:spectrum}, 
we show the spectra of antiproton kinetic energy and photon energy from 
the annihilation $\chi\chi\to 2\varphi \to 2(\bar q q)$ of a 500 GeV DM particle 
with three different mediator masses, $m_{\varphi}=5$, 10 and 50 GeV, respectively.
For a comparison, the spectrum for DM direct annihilation $\chi\chi\to \bar q q$
for the same DM mass is also shown.
For a very light mediator $m_{\varphi}=5$ GeV,  
the antiproton spectrum appears to be a narrow bump.  
As the mediator mass increases, the spectrum becomes broader. 
For $m_{\varphi}=50$ GeV, 
the spectrum looks similar to that from DM direct annihilation,
but  with lower  multiplicity.
Similar observations hold for the photon energy spectrum.
The figure illustrates  that 
the energy spectra of antiprotons and photons  from 
DM annihilation can be highly model dependent, 
which will drastically  change the interpretations of the experimental data.
While the predictions from Pythia for 
low energy hadronic processes have not been fully validated
\cite{Muller:2013BaBar},
the presence of the narrow-bump spectrum is a generic kinematical effect, and 
is insensitive to the details of hardronization model.

\paragraph*{SNR contributions.}
It has been suggested that extra secondary antiprotons can be generated from 
the collisions of primary CRs onto the gas inside the SNRs, and
accelerated by the shock wave in the same way as that of the primary CRs
\cite{
Blasi:2009bd,%
Mertsch:2009ph,%
Berezhko:2014yda,%
Kachelriess:2015oua%
}.
This mechanism does not require 
new class of sources and 
predicts strong correlations between 
the spectra of  secondary species  such as positrons, antiprotons and 
heavy nuclei.
We consider the setup 
in the rest-frame of the shock front (at $x=0$) where 
$u_{1}(u_{2})$ and $n_{1}(n_{2})$ are the upstream (downstream) plasma speed and density respectively.
The compression factor $r=u_{1}/u_{2}=n_{2}/n_{1}$ determines the 
power law index $\gamma=3r/(r-1)$ of  the primary proton spectrum.
Since the secondary antiprotons are produced  inside SNRs
and  propagate  in the same ways as primary protons, 
which largely  cancels the effect of propagation.   
The $\pbar/p$ flux ratio from solving the transportation equation
for $\pbar$ inside SNRs is given by
\cite{
Blasi:2009bd%
}
\begin{align}\label{eq:DSA-solution}
\left(\frac{\Phi_{\pbar}}{\Phi_{p}}\right)_{\text{SNR}}
\sim 
2 n_1 c 
&\left[\gamma
\left(\frac{1}{\xi} +r^2\right)
\int_{m_{p}}^{\mathcal{E}} d\omega \omega^{\gamma-3} \frac{D_1(\omega)}{u_1^2} I(\omega) 
\right.
\nonumber\\
& \left.
+\frac{\tau_{\text{SNR}} \ r}{2 \mathcal{E}^{2-\gamma}} I(\mathcal{E})
\right],
\end{align}
where the first (second) term in the square brackets corresponds to 
the generation of $\pbar$ with (without) acceleration.
The source function $I$ is defined as
$I(\omega)=\int^{\mathcal{E}_{\text{max}}}_{\omega}d\varepsilon \varepsilon^{2-\gamma}\sigma(\varepsilon,\omega)$,
where $\sigma(\varepsilon,\omega)$ is the $\pbar$ production  cross section 
for the process $p(\varepsilon)+\text{H}\to \bar p(\omega)+X$, 
which is taken from \cite{Tan:1982nc}.
The parameter $\xi$ is the fraction of proton energy carried away by a secondary antiproton, and $\tau_{\text{SNR}}$ is  the typical SNR age.
The diffusion coefficient upstream is 
$D_{1}(\mathcal{E})\simeq 3.3\times 10^{22}\mathcal{F}^{-1} (\mathcal{E}/\text{GeV})(B/\mu\text{G})^{-1} \text{cm}^{2}\text{s}^{-1}$.
In the numerical calculation, we fix the parameters as
$u_{1}=0.5\times 10^{8}~\text{cm}\cdot\text{s}^{-1}$,
$n_{1}=2~\text{cm}^{-3}$,
$r=3.8$,
$\xi=0.17$, and
$\tau_{\text{SNR}}=2\times 10^{4}~\text{yr}$.
For the parameters in the diffusion coefficients, we fix them as
$\mathcal{F}=1/20$ and $B=1~\mu\text{G}$.
Note that the normalization of the $\pbar/p$ ratio is proportional to the 
combination $N_{\text{SNR}}=n_{1}u_{1}^{-2}B_{\mu G}^{-1}\mathcal{F}^{-1}$.
There are some uncertainties in the cut-off energy $\mathcal{E}_{\text{max}}$
which depends on the typical SNR age,
but  the typical value is
$\mathcal{E}_{\text{max}}\approx \mathcal{O}(10-100)~\text{TeV}$
\cite{
Blasi:2009bd,%
Mertsch:2014poa%
}.
From \eq{eq:DSA-solution}, one can see that the $\pbar/p$ flux ratio
increases monotonously with increasing energy $E$ and 
saturates when $\mathcal{E}\approx \mathcal{E}_{\text{max}}$.
Thus the generic prediction of the model is a flattening and eventually 
a rise of the $\bar p/p$ ratio in the 100 GeV --  multi-TeV region.

\paragraph*{Fit to AMS data.}
We compare the above-mentioned  three scenarios of antiproton production:
A) DM annihilation into quarks through light  mediators,
B) DM direct annihilation into quarks, and
C) antiprotons produced from the inside of SNRs
through fitting to the high energy AMS-02 antiproton data and 
examine whether they are favoured and can be  distinguished by
the current and  future experiments.
The propagation of  the CR antiprotons is calculated 
using GALPROP v54
\cite{astro-ph/9807150}.
We  consider  three representative  propagation models  selected from 
a global Bayesian analysis to  
the AMS-02 proton and B/C data  using the   GALPROP code
\cite{
Jin:2014ica%
}.
They are selected  to represent the typically
minimal (MIN), median (MED) and maximal (MAX) antiproton fluxes at $95\%$~C.L..
Note that these propagation models  are
different from and  complementary to that given in Ref.  
\cite{Donato:2003xg} which are based on semi-analytical solutions 
with simplified assumptions. %
The Einasto DM profile 
is adopted as a benchmark profile
with a local DM density of $\rho_{0}=0.43\text{ GeV}\text{ cm}^{-3}$.
The effect of solar modulation is taken into account using 
the force-field approximation \cite{Gleeson:1968zza}.
We use charge asymmetric Fisk potentials $\phi_{p}=550$ MV
and $\phi_{\pbar}/\phi_{p}=0.2$ which leads to a good agreement 
with the low-energy antiproton data
\cite{
Hooper:2014ysa%
}.

We determine the model parameters through fitting to the  AMS-02  data. 
Since we are only interested in
the high energy part of the $\pbar/p$  spectrum,
only the $\pbar/p$ data with $E\geq 20$~GeV 
(in total  26 data points) are included,
which also largely avoids the uncertainties in modelling the solar modulation.
Discussions on a possible $\pbar$ excess at low energies can be found in Refs.
\cite{
Hooper:2014ysa,%
Jin:2015sqa,%
Cui:2016ppb%
}.
The significance of an excess over the background  is estimated using 
a test statistics  
$\text{TS}=-2\ln (\mathcal{L}_{\text{bg}}/\mathcal{L}_{\text{bg+src}})$,
where $\mathcal{L}_{\text{bg}}$ and $\mathcal{L}_{\text{bg+src}}$ are 
the likelihood functions for
the scenarios of background-only and 
background plus extra sources, respectively.
In all the fits,
the normalization of  the secondary background is allowed  to 
float freely, i.e. $\Phi_{\pbar,bg}\to\kappa \Phi_{\pbar,bg}$ with 
the normalization constant $\kappa$ determined solely by data.

\begin{figure}[thbp]
\includegraphics[width=0.8\columnwidth]{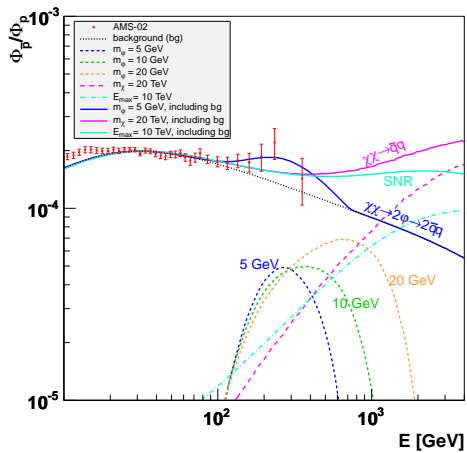}
\caption{
Best-fit $\pbar/p$ flux ratios from the three scenarios:
scenario A with mediator mass $m_{\varphi}=5$, 10 and 20 GeV, respectively,
scenario B with $m_{\chi}$=20 TeV, and 
scenario C with $\mathcal{E}_{\text{max}}=10$ TeV.
The sum with the background in each scenario 
( for $m_{\varphi}=5$ GeV in scenario A) is also shown,
together with the AMS-02 data
\cite{Aguilar:2016kjl}.
}
\label{fig:best-fit-spectra-MED}
\end{figure}

The fit to the data in the background-only scenario in 
the MED propagation model results in 
$\chi^{2}=22.8$ for 25 degrees-of-freedom(d.o.f.),
suggesting a good agreement with the data.
The best-fit background of $\pbar/p$ ratio is 
shown in  \fig{fig:best-fit-spectra-MED}.
Despite the over agreement,
a hint of systematic deviation at energies 
above $\sim 100$ GeV can be seen clearly. 
The best-fit backgrounds  in the MIN and MAX models are found to be  
very close to that in the MED model.

In the scenario  A, i.e.,
$\chi\chi\to 2\varphi\to 2\bar q q$,
we first consider the case of a 5 GeV mediator. %
The best-fit parameters, $\chi^{2}$ and TS values  are 
summarized in \tab{tab:fit-med5GeV}.
In the three propagation models, 
the favoured DM masses are in the range 600 GeV -- 1 TeV.
In the MED model, the best-fit cross section is compatible with
the typical thermal cross section.
From the MIN to the MAX model, 
the variation of the best-fit cross section 
is within an order of magnitude,
which represents the typical uncertainties due to  propagation models. 
Since the measured $\pbar/p$ is very small and of  $\mathcal{O}(10^{-4})$, 
it is a good approximation that 
the background and the new source contributions can be
summed together in $\pbar/p$ flux ratio, i.e.,
$\Phi_{\pbar}/\Phi_{p}\approx 
(\Phi_{\pbar,\text{bg}}/\Phi_{p})+(\Phi_{\pbar,\text{src}}/\Phi_{p})$.
In \fig{fig:best-fit-spectra-MED},
the best-fit $\bar p/p$ flux ratios together with the background
in the MED model are shown.
It can be seen that the structure in the AMS-02 data at 
$\sim 300$ GeV can be  reproduced for $m_{\varphi}=5$~GeV.
For heavier mediators, we find 
$m_{\chi}=1.49  (2.74)$ TeV, $\sigmav=9.4 (23.2)\times 10^{-26}\text{ cm}^{3}\text{s}^{-1}$ and $\chi^{2}=14.3 (14.9)$ for
$m_{\varphi}=10  (20)$ GeV,
and the peaks of the best-fit spectra move to higher energies,
which worsens the agreement with the data, 
and results in larger $\chi^{2}$ values,
as can be seen from  \fig{fig:best-fit-spectra-MED}.

\begin{table}[htbp]
	\centering
		\begin{tabular}{c|c|ccccc}
		\hline\hline
		& Model &  $m_\chi$[GeV] & $\langle\sigma v\rangle(\eta) $ & $\kappa$ & $\chi^2$ & TS \\
		\hline
		\multirow{3}{*}{A } & MIN & $765_{-153}^{+167}$ & $18.6_{-8.0}^{+10.7}$ & $1.12{\pm 0.01}$ &
12.5 &11.6\\
		& MED & $808_{-165}^{+184}$ & $5.18_{-2.37}^{+3.04}$ & $1.13{\pm 0.01}$ &
13.8 & 9.0 \\
		 & MAX & $826_{-168}^{+185}$ & $2.29_{-1.06}^{+1.31}$ & $1.13{\pm 0.01}$ &
15.5 &8.5\\
		\hline
		\multirow{3}{*}{B } & MIN & $20000$ & $1200{\pm 410}$ & $1.12{\pm 0.01}$ &
15.5 &8.6\\
		 & MED & $20000$ & $291{\pm 123}$ & $1.13{\pm 0.01}$ &
17.2 &5.6\\
		 & MAX & $20000$ & $117{\pm 54}$ & $1.12{\pm 0.01}$ &
19.3 &4.7\\
		\hline
		\multirow{3}{*}{C } & MIN & -- & $(0.262{\pm 0.103})$ & $1.08{\pm 0.02}$ &
17.6 &6.5\\
		 & MED &-- & $(0.195{\pm 0.104})$ & $1.10{\pm 0.02}$ &
19.2 &3.5\\
		 & MAX &  -- & $(0.172_{-0.105}^{+0.104})$ & $1.10{\pm 0.02}$ &
21.4 &2.7\\
		\hline\hline
		\end{tabular}
	\caption{ 
 Fit results for the three scenarios:  
 scenario A  %
 with  $m_{\varphi}=5.0$~GeV,	
 scenario B %
 with $m_{\chi}=20$~TeV,	
 and 
 scenario C %
 with  $\mathcal{E}_{\text{max}}=10$~TeV,  in the MIN, MED and MAX propagation models.
The cross section is in units of $10^{-26}\text{cm}^3\text{s}^{-1}$.
The numbers in the parentheses  in the scenario C
stand for the values of the  factor $\eta$.
}
\label{tab:fit-med5GeV}
\end{table}

In the scenario B, i.e., $\chi\chi\to \bar q q$,
we find that the current data impose an lower limit of $m_{\chi}\gtrsim 2$ TeV,
confirming the previous analysis based on the preliminary data
\cite{
Jin:2015sqa%
}.
Introducing a heavier DM mass always improves the agreement with the data. 
However, %
after  $m_{\chi}\gtrsim 10$~TeV,
the value of $\chi^{2}$ gradually ceases to  decrease and 
approaches a constant.
The reason is that for very heavy DM particles, 
only the low-energy tail of the DM generated 
antiproton spectrum can extend to
the region accessible to  the current AMS-02 experiment  ($E\leq 450$ GeV).
In this region, the increase of the DM mass leads to lower $\pbar$ flux 
which  can be compensated by the  increase of the annihilation cross section. 
In \tab{tab:fit-med5GeV}, the fit results for fixed $m_{\chi}=20$~TeV are shown. 
The corresponding best-fit  $\bar p/p$ ratio is shown in \fig{fig:best-fit-spectra-MED}. 

For the scenario C, 
we use the expression of \eq{eq:DSA-solution} plus 
a background contribution.
Besides the normalization factor $\kappa$ of the background, 
the normalization 
of SNR contribution is also  allowed to vary freely by
introducing a factor $\eta$,
i.e., $N_{\text{SNR}}\to \eta N_{\text{SNR}}$.
The spectral shape of the SNR antiprotons is characterized  by
the maximal energy $\mathcal{E}_{\text{max}}$.
Similar to the case of the DM direct annihilation,
we find that a SNR contribution with sufficiently large 
$\mathcal{E}_{\text{max}}\sim \mathcal{O}(10)$~TeV
can improve the agreement with the data, and
the improvement gradually saturates when $\mathcal{E}_{\text{max}}\gtrsim 10$~TeV,
which is due to a similar degeneracy between $\mathcal{E}_{\text{max}}$ and the normalization factor $\eta$. 
We thus fix the cut-off $\mathcal{E}_{\text{max}}$ to be 10 TeV.
The predicted spectrum of $\pbar/p$ ratio in the multi-TeV region is again a
flattening and weak rise until the maximum energy $\mathcal{E}_{\text{max}}$ is reached.

The TS values of the three scenarios in 
the three propagation models are listed in \tab{tab:fit-med5GeV}.
It is clear that the scenario A has the highest significance 
in all the propagation models.
In this scenario, 
as can be seen from  \fig{fig:best-fit-spectra-MED}, 
above 450 GeV,
the $\pbar/p$ ratio is expected to fall with energy,
while in the other two scenarios,
the spectra will continue to rise to higher energies.
Thus this scenario can be distinguished easily from the other two
by the future data. 

\begin{figure}[thbp]
\includegraphics[width=0.8\columnwidth]{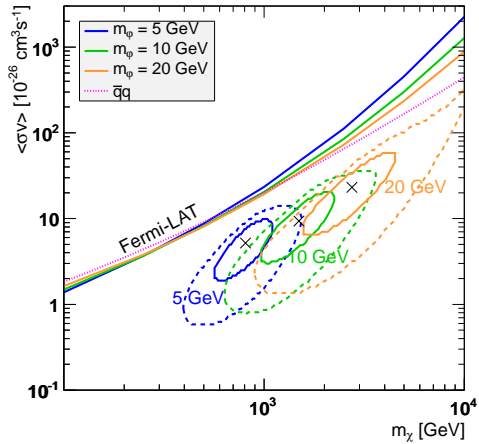}
\caption{
Contours of favoured regions  by the AMS-02 data at $68\%$ (solid) and $95\%$ (dashed) C.Ls. in ($m_{\chi}$, $\sigmav$) plane
in the scenario A %
with  $m_{\varphi}=5$, 10 and 20 GeV, respectively. %
The crosses indicate the best-fit values. 
The $95\%$ C.L. upper limits derived  from the 
Fermi-LAT data on the dSphs are shown as the solid curves.
The derived  limits  for the scenario B are also shown.
}
\label{fig:sumcont_Einasto_MED}
\end{figure}

\paragraph*{Constraints.}
The most stringent and robust constraints  so far on 
the DM annihilation cross sections arise  from 
the observation of $\gamma$-rays towards 
dSphs by the Fermi-LAT collaboration
\cite{
Ackermann:2015zua%
}.
Due to the significant difference in the spectra shape, 
the reported Fermi-LAT limits do not apply to 
the case where DM annihilates through mediators.
We derive the upper limits directly from the likelihood profile per energy bin 
of gamma-rays provided by the Fermi-LAT collaboration
for a selection of 15 dSphs with high-confidence J-factors and backgrounds
\cite{
Carpenter:2016thc%
}.
As a cross check, in \fig{fig:sumcont_Einasto_MED} 
we  show the derived upper limits for 
the DM direct annihilation $\chi\chi\to \bar q q $  which
well reproduces  result of  Fermi-LAT 
\cite{
Ackermann:2015zua%
}.
The limits for  the case with  $m_{\varphi}=5$, 10 and 20~GeV are  shown in  \fig{fig:sumcont_Einasto_MED}, 
together with the regions allowed by the AMS-02 data. %
The favoured regions are all consistent with the current limits.
The derived upper limits turn out to be dependent on 
the mediator mass at high DM mass region.
Compared with the direct DM annihilation,
we find that at high energies around 10 TeV,
the derived upper limit is  weaker by a factor of five.
This is due to the fact that
for a multi-TeV DM particle, 
the relatively narrow $\gamma$-ray  spectrum shown in \fig{fig:spectrum}
has smaller fraction of 
photons entering into the low energy region accessible to Fermi-LAT experiment, 
which results in less stringent constraints.

In the scenario of DM annihilation into light quarks through mediators,
we find that the predicted positron fraction is at most 
$4\times 10^{-4}$ for $m_{\varphi}=5$ GeV, which is far below the expected 
background $\sim \mathcal{O}(10^{-2})$
\cite{
Jin:2013nta,%
Jin:2014ica%
}. 
Thus the constraints from the positron 
fraction is rather weak.
The current LHC search for mono-X plus missing  energy can 
only impose constraints on the mediator mass in the case 
of $m_{\varphi}>2m_{\chi}$
\cite{mono-X},
which makes it less relevant to the case of DM annihilation with 
light mediators.

\paragraph*{Acknowlegements.}
YLW is grateful to  S. Ting for warm hospitality and 
 discussions during his visit to the AMS-02 collaboration at CERN.
This work is supported 
by
the NSFC  under Grants
No.~11335012 and
No.~11475237.

\bibliographystyle{apsrev} %
\bibliography{misc,pbar}
\end{document}